%% file: main.tex
\documentclass[sigconf, nonacm]{acmart}

\providecommand{\compilemode}{1}

\input{style/preamble}

\AtBeginDocument{%
  }

\newcommand\vldbdoi{XX.XX/XXX.XX}
\newcommand\vldbpages{XXX-XXX}
\newcommand\vldbvolume{19}
\newcommand\vldbissue{12}
\newcommand\vldbyear{2026}
\newcommand\vldbauthors{\authors}
\newcommand\vldbtitle{\shorttitle}
\newcommand\vldbavailabilityurl{https://github.com/postechdblab/CADENZA}
\newcommand\vldbpagestyle{empty}

\begin{document}

\ifnum\compilemode<2

\title{CADENZA in Action: Breaking the Monolith with Intent-Dependent Plan Spaces for Semantic Queries}

\author{Jaehyun Ha}
\email{jhha@dblab.postech.ac.kr}
\affiliation{
  \institution{GSAI, POSTECH}
  \city{Pohang}
  \country{Korea}
}

\author{Yongjoo Park}
\email{yongjoo@illinois.edu}
\affiliation{
  \institution{Univ.~of Illinois Urbana-Champaign}
  \city{Urbana}
  \country{USA}
}

\author{Wook-Shin Han}
\authornote{Corresponding author.}
\email{wshan@dblab.postech.ac.kr}
\affiliation{
  \institution{GSAI, POSTECH}
  \city{Pohang}
  \country{Korea}
}

\input{style/macros}

\begin{abstract}
Semantic query processing engines execute semantic operators, whose behavior is specified by natural-language intents, via model inference over multimodal data. Most existing optimizers optimize the operators at the granularity of monolithic implementations---such as LLMs and embedding models---forcing a trade-off between expensive model calls and cheaper alternatives that fail to capture intent-dependent semantics. We present \OURSYSTEM{}, a semantic operator optimizer that compiles an intent into decomposed steps, selects concrete physical implementations for each step, and tunes their parameters under user-specified quality--latency--cost preferences. In this demonstration, users interact with \OURSYSTEM{} through a web interface over multimodal databases, exploring how an intent is decomposed into alternative plans, how each plan is optimized, and how different preferences yield different winning plans.
\end{abstract}

\maketitle

\pagestyle{\vldbpagestyle}
\begingroup\small\noindent\raggedright\textbf{PVLDB Reference Format:}\\
\vldbauthors. \vldbtitle. PVLDB, \vldbvolume(\vldbissue): \vldbpages, \vldbyear.\\
\href{https://doi.org/\vldbdoi}{doi:\vldbdoi}
\endgroup
\begingroup
\renewcommand\thefootnote{}\footnote{\noindent
This work is licensed under the Creative Commons BY-NC-ND 4.0 International License. Visit \url{https://creativecommons.org/licenses/by-nc-nd/4.0/} to view a copy of this license. For any use beyond those covered by this license, obtain permission by emailing \href{mailto:info@vldb.org}{info@vldb.org}. Copyright is held by the owner/author(s). Publication rights licensed to the VLDB Endowment. \\
\raggedright Proceedings of the VLDB Endowment, Vol. \vldbvolume, No. \vldbissue\ %
ISSN 2150-8097. \\
\href{https://doi.org/\vldbdoi}{doi:\vldbdoi} \\
}\addtocounter{footnote}{-1}\endgroup

\ifdefempty{\vldbavailabilityurl}{}{
\vspace{.3cm}
\begingroup\small\noindent\raggedright\textbf{PVLDB Artifact Availability:}\\
The source code, data, and/or other artifacts have been made available at \url{\vldbavailabilityurl}.
\endgroup
}

\input{sections/1_introduction}
\input{sections/2_system_overview}

\input{sections/3_demonstration}

\input{sections/4_conclusion}

\begin{acks}
\camera{This work was partly supported by the National Research Foundation of Korea~(NRF) grant funded by the Korea government~(MSIT) (No.~RS-2025-00517736, 95\%) and the Institute of Information \& Communications Technology Planning \& Evaluation~(IITP) grant funded by the Korea government~(MSIT) (No.~RS-2024-00509258, Global AI Frontier Lab, 5\%).}
\end{acks}

\bibliographystyle{ACM-Reference-Format}
\bibliography{bib/refs}

\fi

\ifnum\compilemode>0

\newpage
\appendix

\ifnum\compilemode=2
    \setcounter{page}{1}
\fi

\fi

\end{document}

%% file: style/preamble.tex
\usepackage[utf8]{inputenc}   
\usepackage{graphicx}
\usepackage{subcaption}
\usepackage{float}  
\usepackage[export]{adjustbox}
\usepackage{multirow}
\usepackage{makecell}
\usepackage{tabularx}
\usepackage{booktabs}
\usepackage[skip=10pt]{caption}
\usepackage{tikz}
\usetikzlibrary{calc}
\usepackage{xcolor}
\usepackage{xspace}
\usepackage{listings}
\usepackage{enumitem}
\usepackage{amsmath}
\usepackage{amssymb}
\usepackage[capitalise,noabbrev]{cleveref}
\usepackage{placeins}

\definecolor{myred}{RGB}{200,0,0}

%% file: style/macros.tex
\newcommand{\camera}[1]{#1}
\newcommand{\wshan}[1]{\textcolor{red}{[#1]}}
\newcommand{\jhha}[1]{\textcolor{red}{#1}}
\newcommand{\yjpark}[1]{\textcolor{teal}{[#1]}}
\newcommand{\OURSYSTEM}[1]{\texttt{CADENZA}\xspace}
\newcommand{\UPQUAL}[1]{0.40}
\newcommand{\UPCOST}[1]{37.2}
\newcommand{\UPLAT}[1]{11.2}
\newcommand{\para}[1]{\medskip\noindent\textbf{#1}}
\newcommand*\circled[1]{\tikz[baseline=(char.base)]{
    \node[shape=circle, fill=black, text=white, draw=black, inner sep=0.5pt] (char) {#1};}}
\newcommand{\TxRA}[1]{\texttt{TxRA}}

\newcommand{\map}[1]{\texttt{Map}}
\newcommand{\filter}[1]{\texttt{Filter}}
\newcommand{\join}[1]{\texttt{Join}}
\newcommand{\groupby}[1]{\texttt{GroupBy}}
\newcommand{\topk}[1]{\texttt{Top-K}}

\newcommand{\semfilter}[1]{\textsf{SemFilter}}

\newcommand{\semop}[1]{$SO$}
\newcommand{\groundtruth}[1]{$GT_{D_s}$}
\newcommand{\samples}[1]{$D_s$}
\newcommand{\samplesize}[1]{$|D_s|$}
\newcommand{\physicalplan}[1]{$P_P$}
\newcommand{\dataset}[1]{$D$}

\newcommand{\BIODEXMAXLAT}[1]{9.87}
\newcommand{\BIODEXMAXCOST}[1]{152}
\newcommand{\BIODEXMAXQUAL}[1]{0.06}

\newcommand{\TEXTEMBEDDINGSMALL}[1]{\texttt{Text-Embedding-3-Small}}
\newcommand{\GPTFOUR}[1]{\texttt{GPT-4.1}}
\newcommand{\GPTFOURMINI}[1]{\texttt{GPT-4.1-mini}}
\newcommand{\GPTFOURNANO}[1]{\texttt{GPT-4.1-nano}}

\newcommand{\SIMILARITYFILTER}[1]{\texttt{Similarity Filtering}}
\newcommand{\DIRECTGPT}[1]{\texttt{Direct GPT-4.1}}
\newcommand{\DIRECTGPTMINI}[1]{\texttt{Direct GPT-4.1 Mini}}
\newcommand{\DIRECTGPTHOL}[1]{\texttt{Direct GPT-4.1 (Holistic)}}
\newcommand{\DIRECTGPTMINIHOL}[1]{\texttt{Direct GPT-4.1 Mini (Holistic)}}
\newcommand{\DIRECTGPTITER}[1]{\texttt{Direct GPT-4.1 (Iterative)}}
\newcommand{\DIRECTGPTMINIITER}[1]{\texttt{Direct GPT-4.1 Mini (Iterative)}}
\newcommand{\CONFIDENCECASCADE}[1]{\texttt{Confidence Cascades}}
\newcommand{\SIMILARITYCASCADE}[1]{\texttt{Similarity Cascades}}
\newcommand{\ITERATIVEGENERATION}[1]{\texttt{Iterative Generation}}

%% file: sections/1_introduction.tex
\section{Introduction}
\label{sec:introduction}

\begin{figure}[t]
  \centering
  \includegraphics[width=\columnwidth]{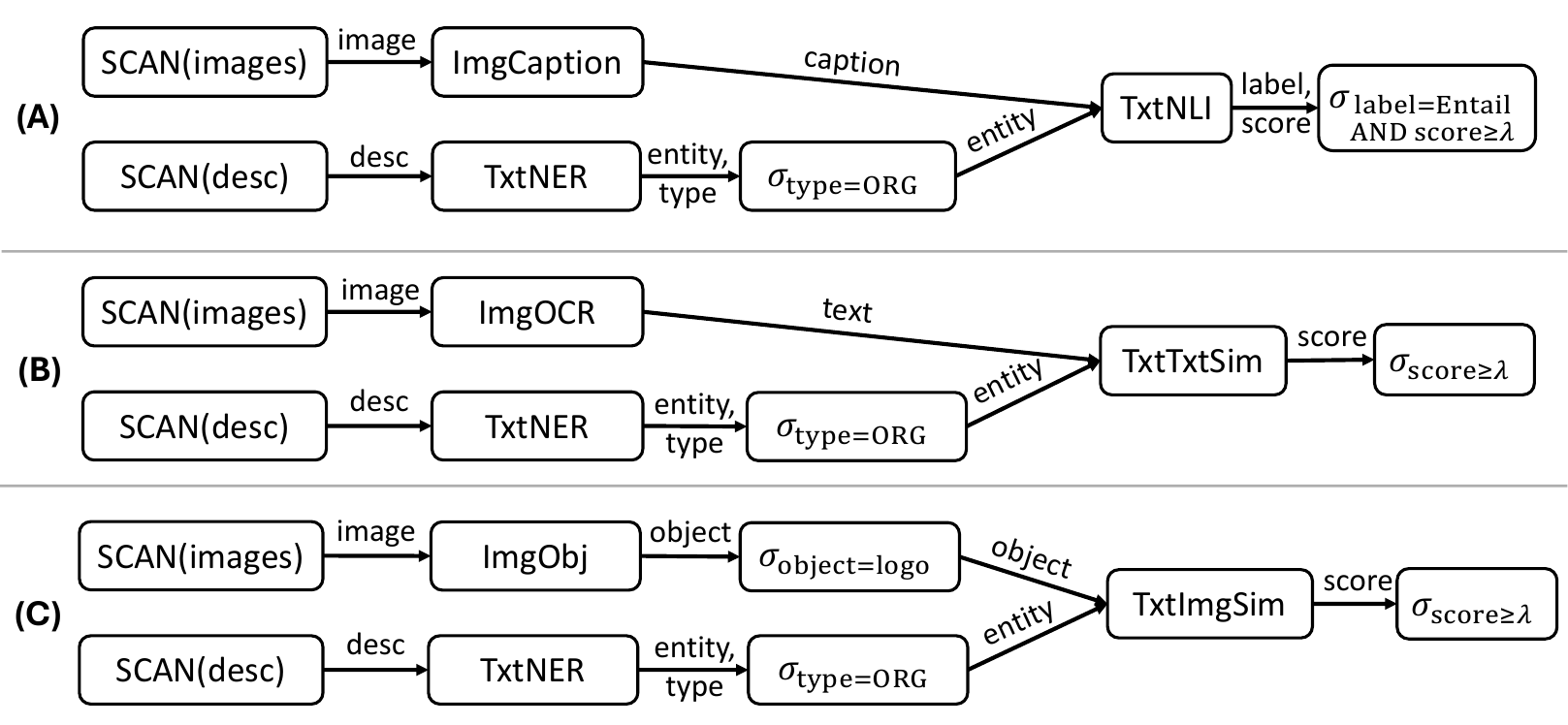}
  \caption{Three candidate logical plans for \texttt{SemJoin} with \emph{``match product descriptions with product photos of the same brand.''} Each node is an operator and each edge denotes the attribute consumed by the next operator.}
  \label{fig:logical_plans}
\end{figure}

With the advent of Large Language Models (LLMs), a new class of \emph{semantic query processing engines} (SQPEs) has emerged~\cite{LOTUS, abacus, AOP, unify, palimpzest}. Semantic operators---such as \texttt{SemFilter}, \texttt{SemJoin}, and \texttt{SemMap}---are operators whose behaviors are specified by natural-language task descriptions (i.e., intents) and executed via model inference over multimodal data (e.g., text and images). These operators simplify complex multimodal analysis. For example, suppose an e-commerce platform onboards a vendor who delivers product descriptions (text with brand, category, specs, etc.) and product photos (images with logos but no metadata) as databases with no shared identifier. To build a unified catalog, a data engineer can execute a \texttt{SemJoin} to \emph{``match product descriptions with product photos of the same brand.''}

Most existing semantic query optimizers~\cite{palimpzest, abacus, LOTUS, AOP, unify, DOCETL} optimize at the granularity of a monolithic operator implementation---e.g., choosing an LLM, an embedding model, or a cascade---rather than exploring alternative decompositions of the operator's task. A large LLM can ensure high accuracy yet incurs substantial cost and latency; a smaller LLM or embedding model reduces cost, but embeddings are often too coarse to capture the nuances of arbitrary intents. For instance, embedding an entire product description into a single vector dilutes fine-grained brand signals among other attributes, degrading match quality regardless of the model chosen. The root cause is that these systems never look inside the intent: they cannot decompose it into finer-grained, specialized steps that would admit both cheaper and more accurate execution.

Our key insight is that an intent can be compiled into a small logical DAG where each node is a specialized model or a control-flow operator, dynamically selected or synthesized in a query-aware manner. This enables both cheaper and more accurate execution than monolithic approaches. For example, the brand-matching query above can be translated into any of a few candidate plans (\cref{fig:logical_plans}). While each plan extracts brand entities via NER, they employ different visual processing strategies: Plan~A generates captions (e.g., via BLIP~\cite{BLIP}) and verifies brand entailment via NLI (e.g., via DeBERTa~\cite{he2021deberta})---cheap, but accuracy depends on whether the caption mentions the brand; Plan~B reads printed text via OCR for direct text comparison---accurate when brand text is visible, but fails otherwise; and Plan~C detects logos (e.g., via YOLO~\cite{YOLO}) and matches them via text--image similarity (e.g., CLIP~\cite{CLIP})---the most robust but also the most expensive. The optimal plan is chosen considering the user's preference over quality, latency, and cost.

We present \OURSYSTEM{}, a semantic operator optimizer that automatically discovers such alternative decompositions, selects concrete implementations for each decomposed step, tunes their parameters under multi-objective optimization, and returns the best plan under the user's quality--latency--cost preference. In this demonstration, users interact with \OURSYSTEM{} through an interactive web interface over multimodal databases, exploring how an intent is decomposed into alternative plans, how each plan is optimized, and how different preferences yield different winning plans.

\begin{figure}[t]
  \centering
  \includegraphics[width=0.95\columnwidth]{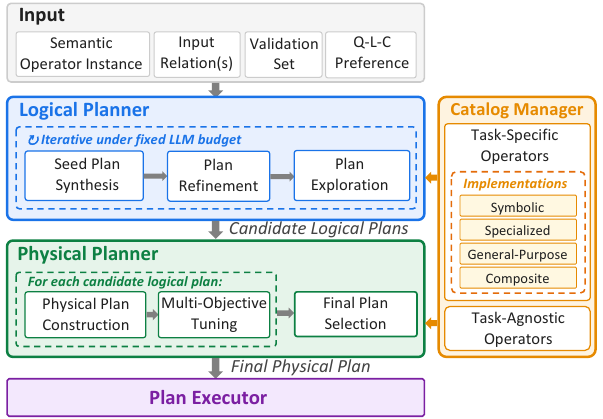}
  \caption{System architecture of \OURSYSTEM{}.}
  \label{fig:architecture}
\end{figure}

\begin{figure*}[t]
  \centering
  \begin{subfigure}[t]{0.33\textwidth}
    \includegraphics[width=\textwidth]{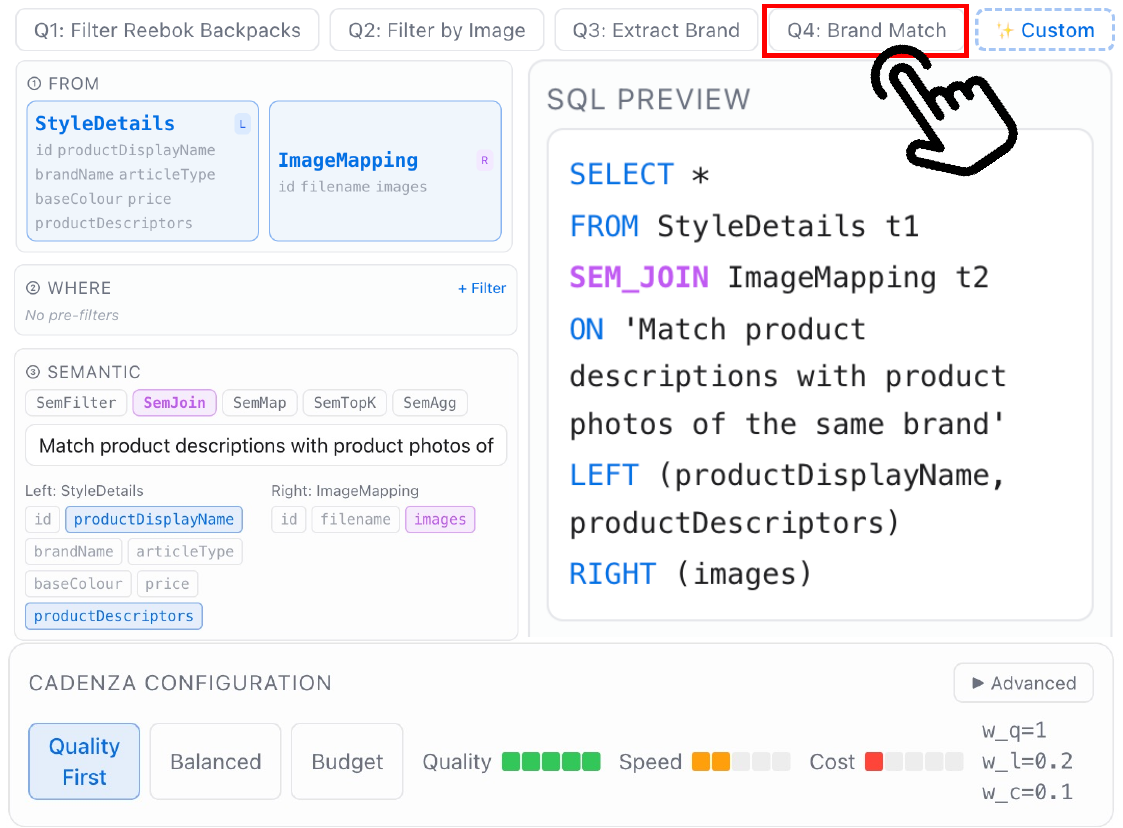}
    \caption{Query builder}
    \label{fig:walk_query}
  \end{subfigure}\hfill
  \begin{subfigure}[t]{0.33\textwidth}
    \includegraphics[width=\textwidth]{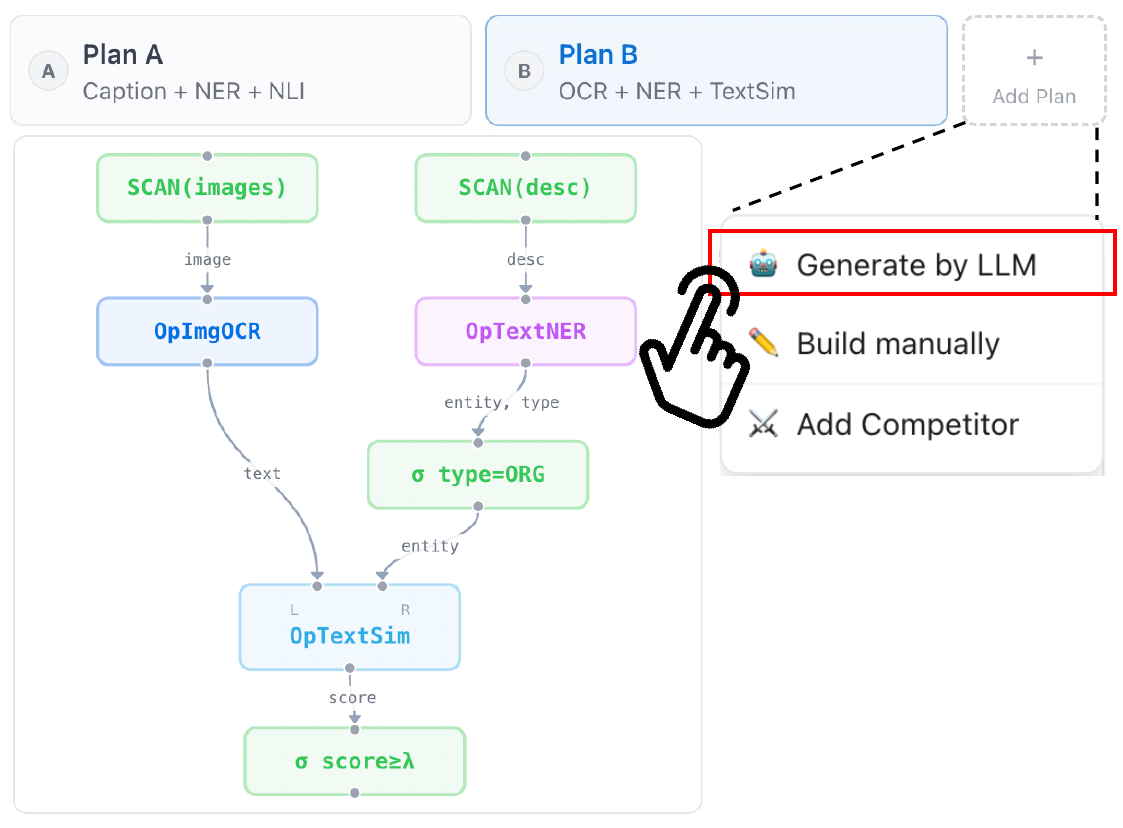}
    \caption{Logical plan generation}
    \label{fig:walk_logical}
  \end{subfigure}\hfill
  \begin{subfigure}[t]{0.33\textwidth}
    \includegraphics[width=\textwidth]{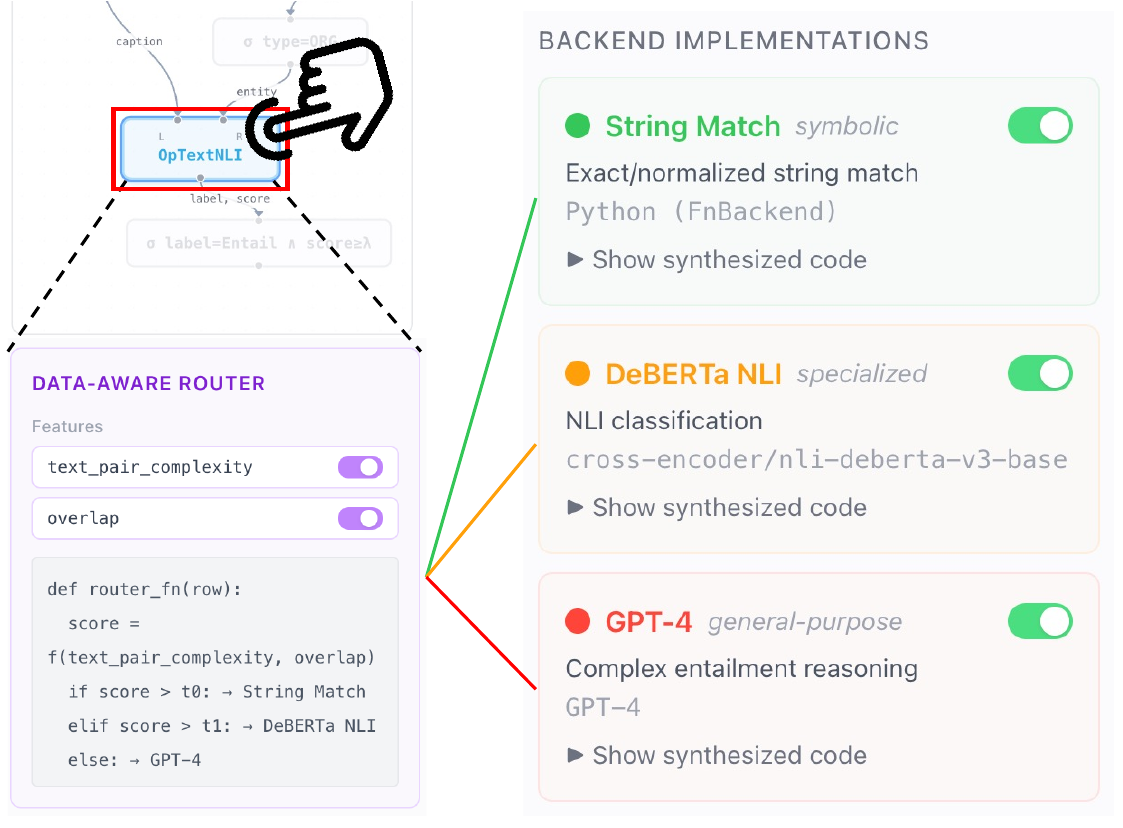}
    \caption{Physical plan construction}
    \label{fig:walk_physical}
  \end{subfigure}

  \vspace{4pt}
  \begin{subfigure}[t]{0.33\textwidth}
    \includegraphics[width=\textwidth]{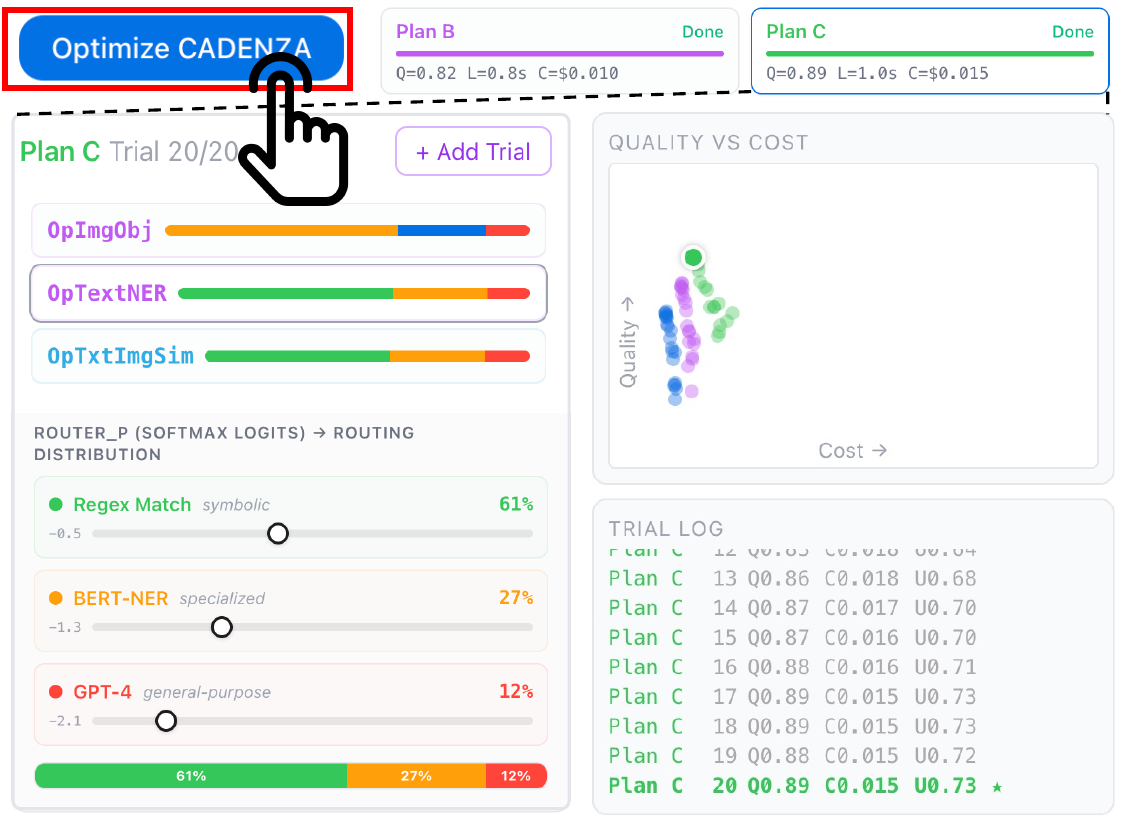}
    \caption{Multi-objective tuning}
    \label{fig:walk_bo}
  \end{subfigure}\hfill
  \begin{subfigure}[t]{0.33\textwidth}
    \includegraphics[width=\textwidth]{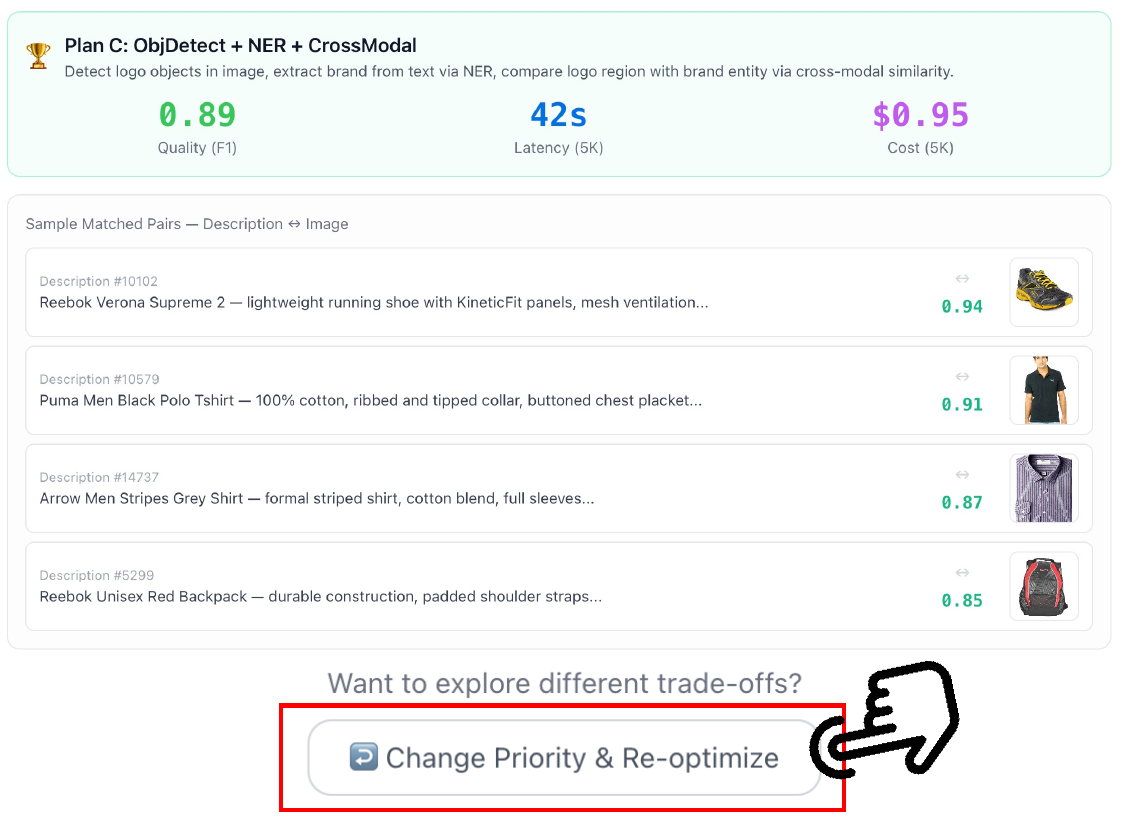}
    \caption{Re-optimization}
    \label{fig:walk_reopt}
  \end{subfigure}\hfill
  \begin{subfigure}[t]{0.33\textwidth}
    \includegraphics[width=\textwidth]{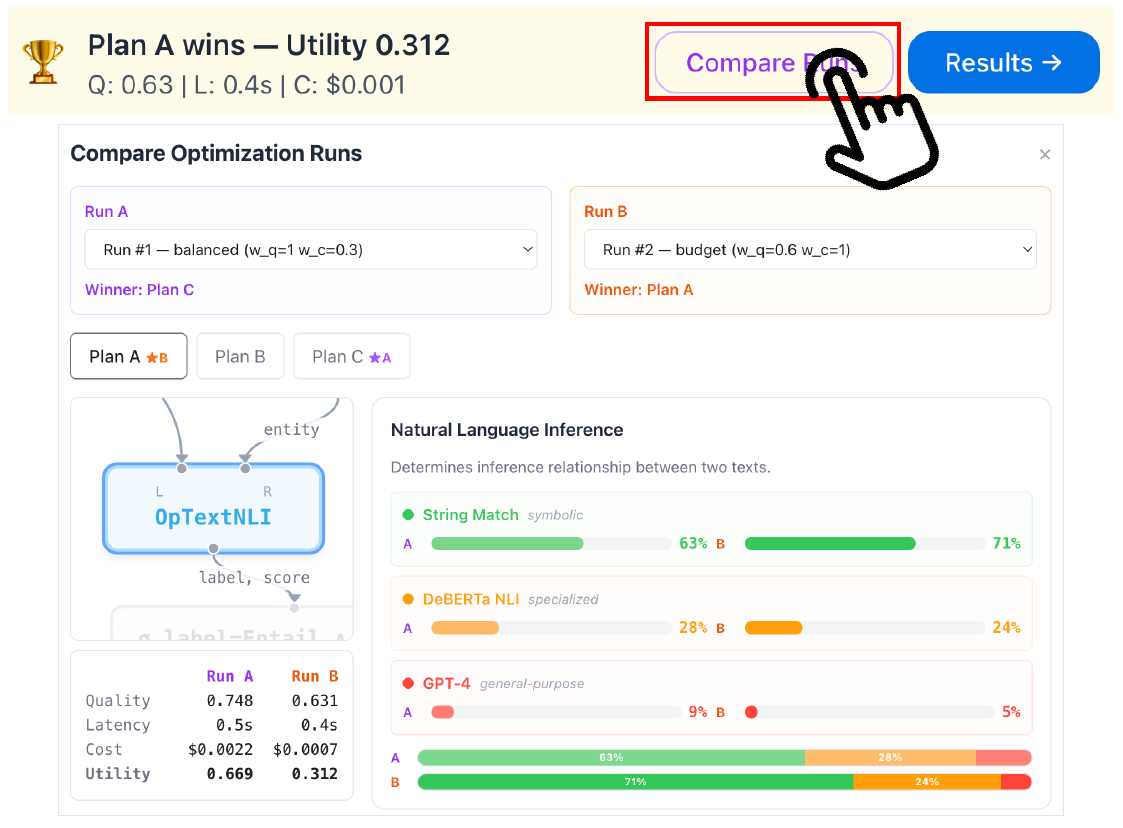}
    \caption{Comparing runs}
    \label{fig:walk_winner}
  \end{subfigure}
  \caption{Demo walkthrough. \textbf{S1}~(a--d): Under Balanced preference, a SemJoin query is decomposed into three plans; Plan~C wins. \textbf{S2}~(e--f): Under Budget preference, re-optimization yields a different winner, Plan~A.}
  \vspace{-2mm}
  \label{fig:demo_walkthrough}
\end{figure*}

%% file: sections/2_system_overview.tex
\section{System Overview}
\label{sec:system_overview}

This section overviews \OURSYSTEM{}, a standalone optimizer with a built-in plan executor. Integration with existing SQPEs is supported through lightweight dispatch hooks and adapter patterns. We describe the system's inputs and operator catalog, then present the logical planner that decomposes intents into logical plans, and the physical planner that compiles, tunes, and selects the final plan under user-specified quality--latency--cost preferences (Figure~\ref{fig:architecture}). 

\para{Input.}
\OURSYSTEM{} takes as input (i)~a \emph{semantic operator} (e.g., \texttt{SemJoin}) and its \emph{intent}, (ii)~the multimodal input relation(s), (iii)~an optional \emph{validation set}, and (iv)~a quality--latency--cost (Q-L-C) \emph{preference}. \camera{Formally, an \emph{intent} is a natural-language specification of a semantic operator instance's desired behavior, with no fixed algebraic form; the same intent therefore admits multiple valid decompositions that \OURSYSTEM{} surfaces as alternative logical plans.} Preferences are weight vectors $(w_q, w_l, w_c)$, with preset profiles \emph{Quality First}, \emph{Balanced}, and \emph{Budget}.
Following prior work~\cite{LOTUS,abacus}, we estimate each candidate plan's quality, latency, and cost on a small \emph{proxy-labeled} validation set. If none is provided, \OURSYSTEM{} labels a subset with an oracle model (e.g., GPT-4), reused across all candidates. \camera{Because proxy labels are used only to \emph{rank} candidate plans rather than to define ground truth, the ranking stays stable under moderate labeling noise; users may also supply their own validation set.} \camera{Note that $U$ scores only the selected plan's deployment cost; the (plan-dependent) optimization overhead is reported separately in the trial log and amortized over deployment rather than charged to $U$.}

\para{Catalog Manager.}
The catalog manager maintains the logical operators and their physical implementations. Operators fall into two classes. \emph{Task-specific operators} are typed, per-tuple inference steps that each correspond to a well-defined task formulation in multimodal model ecosystems (e.g., Hugging Face), with a typed signature and a stable output schema independent of backend. Each runs inference on input attributes and appends produced columns (e.g., \textsf{ImgCap} appends a \textit{Caption} column). We group task-specific operators by the database role of their outputs into five categories: \emph{structural} (indexes substructures, e.g., splitting, region detection), \emph{attributive} (unary descriptors, e.g., classification, NER), \emph{associative} (relations over elements or pairs, e.g., NLI, pair scoring), \emph{generative} (derived artifacts as new semantic objects, e.g., summarization, captioning), and \emph{latent} (vectors for similarity-based retrieval). \emph{Task-agnostic operators}---standard relational operators such as filter ($\sigma$), join ($\bowtie$), and aggregation ($\gamma$)---connect task-specific steps within a plan. The full catalog, spanning text and image modalities, is browsable in the demo interface; new operators are added by registering a task signature and assigning a category.

On the physical side, each operator has one or more \emph{implementations}, organized by backend family: \emph{symbolic} (e.g., string matching), \emph{specialized} (e.g., DeBERTa-NLI, CLIP), \emph{general-purpose} (e.g., GPT-4), and \emph{composite} (e.g., multi-stage pipelines). For the supported operators, we precompile implementations offline, except symbolic ones that require query-specific knowledge and are instantiated on demand. Precompiled implementations are maintained via an agentic pipeline that automatically discovers and integrates models from Hugging Face, keeping the catalog up to date. Task-agnostic operators use standard implementations.

\para{Logical Planner.}
The logical planner determines \emph{what} to compute by decomposing the intent into a DAG of operators---a \emph{logical plan}. Each plan interleaves task-specific operators with task-agnostic operators that wire intermediate results together. Because a single intent admits multiple valid decompositions, the planner first synthesizes a small set of \emph{seed plans} under a fixed LLM budget: it prompts an LLM with the intent, schemas, catalog, and few-shot examples to propose a decomposition. Each draft is verified against schema and dataflow constraints, and invalid drafts are refined (\emph{plan refinement}). The planner then expands the seed set via rule-based transformations (\emph{plan exploration}) without additional LLM calls, producing structurally diverse candidate plans.

Figure~\ref{fig:logical_plans} shows three candidates for our running example; all extract brand entities from descriptions via NER, but differ in how they process images.
Plan~A (\emph{NER+Caption+NLI}) captions each image, then takes the cross product of captions and extracted brand entities, scores entailment via NLI, and filters on the score ($\sigma_{\mathit{score}\ge\lambda}$).
Plan~B (\emph{NER+OCR+TextSim}) reads text printed on the product via OCR and compares it to the brand entity via text similarity.
Plan~C (\emph{NER+ImgObj+TxtImgSim}) detects objects in images, filters for logo regions ($\sigma_{\mathit{object}=\mathit{logo}}$), scores each logo--brand pair via CLIP-based text--image similarity, and filters on the score ($\sigma_{\mathit{score}\ge\lambda}$).

\para{Physical Planner.}
The physical planner determines \emph{how} to execute each logical plan by binding every operator to concrete implementations with tunable parameters. Task-agnostic operators such as $\sigma_{\mathit{sim}\ge\lambda}$ become standard filters with a tunable threshold~$\lambda$. Each task-specific operator is compiled into a \emph{routed ensemble}: the planner retrieves candidate backends from the catalog and synthesizes a \emph{data-aware router} by prompting an LLM with the operator specification and the ordered backend set. The LLM generates (i)~lightweight feature-extraction code that computes cheap, non-inferential signals (e.g., text length, keyword overlap) and (ii)~scoring code that maps these features to a difficulty score $s\in[0,1]$. A fixed bucketing template then uses $s$ and $N\!-\!1$ tunable cutpoints to dispatch each tuple to one of $N$~backends.
For instance, the \textsf{TxtNLI} operator in Plan~A may route pairs whose caption already contains the brand name to a symbolic string matcher, while ambiguous pairs are forwarded to DeBERTa-NLI or GPT-4. The routing thresholds and per-backend parameters together form the tunable space~$\Theta$.

For each candidate plan, \OURSYSTEM{} tunes $\Theta$ via Bayesian optimization on the validation set. Each trial executes the physical plan on the validation set and measures quality~$Q$ (e.g., F1), latency~$L$, and cost~$C$ (API plus local GPU cost). 
By default, these are combined into a scalar utility $U = w_q\,Q - w_l\,\hat{L} - w_c\,\hat{C}$, \camera{where $Q\in[0,1]$ and $\hat{L},\hat{C}$ are latency and cost \emph{log-normalized} to $[0,1]$. The weights $(w_q,w_l,w_c)\ge 0$ thus encode \emph{relative} importance rather than trading raw seconds against dollars, and users set them via the preset profiles or a slider.} Other formulations---such as maximizing one metric subject to hard constraints on the others---are also supported, as the tuning loop is agnostic to the choice of objective function. The best plan is then selected for execution. Crucially, the Q-L-C preference steers not only routing parameters but potentially the winning \emph{logical plan} itself: a \emph{Quality First} setting may favor Plan~C's targeted logo matching, whereas a \emph{Budget} setting may prefer Plan~A's lighter captioning strategy.

%% file: sections/3_demonstration.tex
\section{Demonstration: Two Scenarios}
\label{sec:demonstration}

We demonstrate \OURSYSTEM{}'s end-to-end workflow---from query construction through plan decomposition, physical compilation, and multi-objective tuning---on the E-Commerce scenario from SemBench~\cite{sembench}\footnote{We pre-load the $\text{SF}{=}100$ subset (100 descriptions, 100 images) with cached model weights. Under the default budget, the scenario 1 completes in approximately 120\,s and the scenario 2 in 90\,s.}. The user plays a data engineer who must match product descriptions with product photos of the same brand via \texttt{SemJoin}. During the demo, the user explores how this intent yields different logical plans, how each is compiled and tuned, and how changing preferences selects a different winner.

Beyond the curated scenarios, the interface supports custom dataset registration and flexible plan construction via a drag-and-drop canvas. Users can assemble logical plans from 36 operators---22 for text and 14 for images---while real-time schema validation ensures logical correctness. \camera{For example, a user can register a new image collection and assemble a \texttt{SemFilter} for \emph{``images that display a given phrase''} on the canvas---chaining \textsf{OCR}~$\to$~string match---exercising a different operator and modality from the SemJoin scenario without writing code.}

\subsection{S1: One Intent, Three Strategies}
\label{sec:scenario_a}

\para{(a) Query Builder.}
In our first scenario, the user opens the query builder (\cref{fig:demo_walkthrough}a), selects the E-Commerce dataset and picks preset~Q4, a \texttt{SemJoin} that binds two input relations: product descriptions (text) and product images. The preference panel is set to \emph{Balanced}, and the validation set size and tuning budget are configured through advanced settings.

\para{(b) Logical Plan Generation.}
Clicking \emph{``Generate by LLM''} triggers the logical planner, which synthesizes three candidate plans (\cref{fig:demo_walkthrough}b), each a distinct plan for the same intent (see also \cref{fig:logical_plans}).
By clicking \emph{``Add Competitor,''} the user can also add monolithic baselines for comparison---plans that execute the \texttt{SemJoin} without decomposition. For this walkthrough, the user adds two baselines; their physical implementations are chosen in the next step.

\para{(c) Physical Plan Construction.}
The user now configures physical implementations for each plan (\cref{fig:demo_walkthrough}c). Selecting Plan~A, for example, reveals the routed ensemble for each task-specific operator. The \textsf{TxtNLI} operator lists three candidate backends---a symbolic string matcher, a specialized DeBERTa-NLI model, and GPT-4---each with its synthesized implementation code expandable on click. The user can include or exclude individual backends; removing the symbolic backend shifts the router's dispatch distribution toward the specialized and general-purpose alternatives, a change visible in real time. For operators the user does not explicitly configure, all available backends are included by default and the router is generated automatically.
For the two baselines, the user chooses from several implementation options for the \texttt{SemJoin} operator: a single-LLM plan, an embedding-only plan, or a cascade that tries embeddings first and falls back to an LLM. In this walkthrough, the user selects LLM-based and Embedding-Cascade implementations as the two alternatives to compare against the decomposed plans.

\para{(d) Multi-Objective Tuning.}
The user clicks \emph{``Optimize CADENZA''} to launch Bayesian optimization, which tunes each decomposed plan's routing thresholds and backend parameters on the validation set (\cref{fig:demo_walkthrough}d). A live dashboard tracks progress through a trial log, a Pareto scatter, and per-operator routing cards. Under Balanced preferences, Plan~C's fine-grained decomposition lets the router assign easy inputs to cheap backends while reserving LLM calls for hard cases. The user can also drag a routing-ratio slider to set backend proportions and fire a trial, observing how the metric triple responds.
The user then clicks \emph{``Optimize Baselines''} to tune the two monolithic plans under the same setting; their results join the same Pareto chart; \camera{\cref{tab:qlc} reports the measured numbers.}

\begin{table}[h]
\centering
\caption{\camera{$Q$, $L$, $C$ measured by executing each plan---\OURSYSTEM{} and the two baselines---on the demo's running example.}}
\vspace{-2mm}
\label{tab:qlc}
\camera{%
\begin{tabular}{lrrr}
\toprule
Plan & $Q\uparrow$ & $L$\,(s)\,$\downarrow$ & $C$\,(\$)\,$\downarrow$ \\
\midrule
\textbf{\OURSYSTEM{} (ours)} & \textbf{0.82} & \textbf{566} & \textbf{0.80} \\
Single LLM & 0.76 & 25{,}323 & 2.59 \\
Embedding Cascade & 0.63 & 4{,}208 & 1.20 \\
\bottomrule
\end{tabular}}
\end{table}

\subsection{S2: Preference Changes Plan}
\label{sec:scenario_b}

\para{(e) Re-optimization.}
After tuning completes, the results screen displays Plan~C's quality, latency, and cost alongside sample matched pairs. While Plan~C delivers the best quality under Balanced preferences, its cost may be too high for production at scale. The user clicks \emph{``Change Priority \& Re-optimize''} (\cref{fig:demo_walkthrough}e), switching to \emph{Budget} to re-optimize with a cost-conscious objective.

\para{(f) Comparing Runs.}
Once re-optimization finishes, the user clicks \emph{``Compare''} to open a comparison modal (\cref{fig:demo_walkthrough}f) that places the two optimization runs---Balanced (Plan~C wins) and Budget (Plan~A wins)---side by side. Per-operator routing distributions reveal how the Budget objective shifts dispatch away from expensive backends. Plan~A (NER+Caption+NLI), with its lighter image processing, now wins by relying on cheap captioning, not object detection. Preferences thus steer not only routing parameters but the choice of \emph{logical plan} itself: the same intent yields structurally different optimal strategies depending on the quality--cost trade-off.

%% file: sections/4_conclusion.tex
\section{Conclusion}
\label{sec:conclusion}

We demonstrated \OURSYSTEM{}, a semantic operator optimizer that decomposes an intent into alternative logical plans, compiles each into a physical plan with routed ensembles of heterogeneous backends, and tunes them under user-specified quality--latency--cost preferences. Through an interactive demo, users can explore how an intent is decomposed, how each plan is optimized, and how changing preferences yields a different winning plan---making intent-dependent semantic query optimization transparent and practical.